\documentclass[journal,10pt]{IEEEtran}
\renewcommand{\IEEEbibitemsep}{0pt plus 2pt}
\makeatletter
\IEEEtriggercmd{\reset@font\normalfont\footnotesize}
\makeatother
\IEEEtriggeratref{1}
\usepackage[nocomma]{optidef}
\usepackage{amsmath, amsfonts}
\usepackage{indentfirst}
\usepackage[sorting=none]{biblatex}
\usepackage{graphicx}
\usepackage{caption}
\usepackage{xcolor}
\usepackage{soul}
\usepackage{amsmath}
\usepackage{float}
\usepackage{array}
\usepackage{makecell}
\usepackage{enumitem}
\usepackage{textcase}
\usepackage{algorithm}
\usepackage{algorithmicx} 
\usepackage{algcompatible}
\usepackage[noend]{algpseudocode}
\usepackage{subcaption}

\bibliography {references.bib}

\begin{document}

\author{\large\IEEEauthorblockN{Fatima Haouari,
Emna Baccour, Aiman Erbad, Amr Mohamed, and Mohsen Guizani\\}
\IEEEauthorblockA{CSE department, College of Engineering, Qatar University
}}
\title{QoE-Aware Resource Allocation for Crowdsourced Live Streaming: A Machine Learning Approach}
\maketitle
\begin{abstract}
Driven by the tremendous technological advancement of personal devices and the prevalence of wireless mobile network accesses, the world has witnessed an explosion in crowdsourced live streaming.
Ensuring a better viewers quality of experience (QoE) is the key to maximize the audiences number and increase streaming providers' profits. This can be achieved by advocating a geo-distributed cloud infrastructure to allocate the multimedia resources as close as possible to viewers, in order to minimize the access delay and video stalls. Moreover, allocating the exact needed resources beforehand avoids over-provisioning, which may lead to significant costs by the service providers. In the contrary, under-provisioning might cause significant delays to the viewers. 
In this paper, we introduce a prediction driven resource allocation framework, to maximize the QoE of viewers and minimize the resource allocation cost. First, by exploiting the viewers locations available in our unique dataset, we implement a machine learning model to predict the viewers number near each geo-distributed cloud site. Second, based on the predicted results that showed to be close to the actual values, we formulate an optimization problem to proactively allocate resources at the viewers proximity. Additionally, we will present a trade-off between the video access delay and the cost of resource allocation. 
\end{abstract}

\IEEEpeerreviewmaketitle
\section{Introduction}
Crowdsourced live video streaming is on the rise, and it continues to grow every single day. As per Cisco mobile video traffic statistics, 
mobile video content is predicted to present 82\% of the global Internet traffic in 2021 as opposed to 73\% in 2016 \cite{cisco_2017}.
The rise in popularity of crowdsourced live streaming can be attributed to technological advancement, proliferation of smartphones and wireless network availability, which have led crowdsourcers to broadcast their live videos to various content providers.
One of the most popular live streaming platform is Facebook, which had 2.19 billion active users per month in the first quarter of 2018 \cite{facebookstatista}. As per \cite{wordstream} 78\% of Facebook online users are watching live videos, and 1 out of 5 videos on Facebook is live.

The industry and academia have shown an overwhelming interest in crowdsourced streaming recently in terms of achieving the best QoE as it is the key to increase the audiences number and the content providers' revenues. A series of recent studies have been conducted to determine the main factors that affect the viewers' QoE \cite{balachandran2013developing,krishnan2013video}. These studies revealed that viewers QoE is primarily dependent on two factors: First, the video startup delay and playback buffering stalls and second, the video quality which depends on the viewers' internet connectivity quality and available video representations. The authors in \cite{krishnan2013video} highlighted that the higher the startup delay is, the more the viewers abandonment increases. 
They also showed that viewers who experienced low QoE are less likely to revisit the content provider's application within a specific period of time. Therefore, video startup and rebuffering delays have high impact on viewers' QoE. However, the challenge is to serve the viewers with the best QoE possible, while minimizing the cost of resource allocation.

Geo-distributed clouds are proposed to enhance the QoE. In this context, many efforts are working on presenting an efficient resource allocation by proposing heuristics and optimisations. Wu et al. \cite{wu2015scaling} formulated an optimal viewing request distribution in the geo-distributed clouds, they predicted users future demands based on their social influences using an epidemic model. He et al. \cite{he2016coping} presented a resource allocation framework to allocate geo-distributed cloud service to crowdsourcers for transcoding and serving viewers. K. Bilal et al. \cite{bilal2018qoe} presented a QoE-aware resource allocation optimization for crowdsourced multiview live streaming to choose the optimal transcoding cloud site location, and the optimal set of video representations.
The drawback of these traditional algorithms is the near optimal solutions they provide. They lack the ability to allocate the exact resources needed beforehand. This may either lead to over-provisioning of resources that may incur significant costs to the service providers, or under-provisioning of resources that may cause delays to the viewers. Therefore, addressing such a trade-off proactively is a real challenge that requires some accurate prediction techniques.

In this work, we are addressing the proactive resource allocation by adopting machine learning techniques for designing a predictive model for the viewers' locations. In particular, we consider predicting the number of viewers near each geo-distributed cloud site for each incoming live video, in order to proactively allocate resources at the proximity of the viewers. To the best of our knowledge, there is no research work that applied machine learning techniques for resource allocation to maximize QoE and minimize the cost. Only a few studies adopted machine learning to improve the viewers QoE, with their focus varies from dealing with the buffering and the bitrate selection  \cite{petrangeli2017machine}, to determining Adaptive Bitrate (ABR) best parameters in order to improve adaptive video streaming \cite{le2018improving}. The authors in \cite{petrangeli2017machine} proposed a video freeze predictive model to detect possible factors that lead to video stalling at the viewers side. A recent study by \cite{le2018improving} proposed using decision trees to choose the best ABR parameters to improve the adaptive video streaming.
Moreover, few recent studies have used machine learning for predicting the viewers' QoE. The authors in \cite{zhu2016user} predicted the users engagement score, by considering users engagement as a function of Quality of Service (QoS) factors and viewers preferences. Another work in \cite{balachandran2013developing} proposed a classification model for users engagement, where users engagement was quantified in terms of users number of visits and video watching time.

The contributions of this paper are summarized as:
\begin{itemize}
\item Using our collected Facebook 2018 live videos dataset \cite{facebookvideoslive18} containing records of viewers' locations for each video, we develop a regressive model using machine learning techniques that predicts the number of viewers near different geo-distributed cloud sites for each incoming live video.
\item To serve the predicted viewers such that they experience the minimum startup delay with a minimal cost to the content provider, we formulate an optimization problem for allocating resources as near as possible to the viewers.
\end{itemize}

The rest of this paper is organized as follows:
Section \ref{systemModel} presents our system model composed of: (1) viewers predictive model; (2) proactive resource allocation optimizer. We evaluate our system and present a trade-off between minimizing latency and maximizing cost gain in Section \ref{section:performance}. Finally, section \ref{section:conclusion} concludes the paper and discusses the future directions.
\section{System model}\label{systemModel}
\begin{figure}[h]
\captionsetup{font=scriptsize}
  \includegraphics[width=\linewidth]{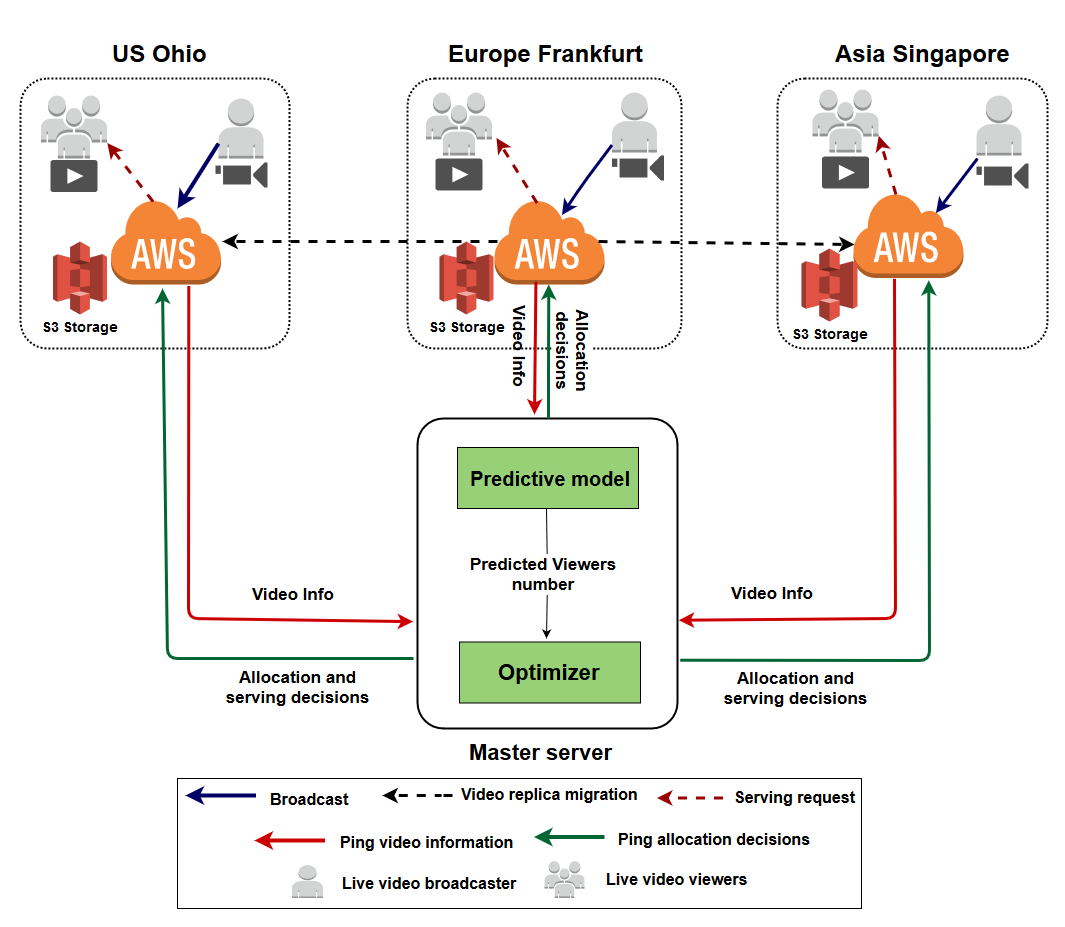}
  \caption{System model.}
  \label{fig:system model}
\end{figure}

In our system, we adopt a geo-distributed cloud infrastructure as shown in Fig. \ref{fig:system model} that consists of multiple geographically distributed cloud sites owned by a content provider. Our predictive model and resource allocation optimizer are deployed in a centralized master server. A set of geo-distributed crowdsourcers broadcast their videos in real time, which will be allocated by default in their nearest cloud site. Each broadcaster cloud site will report the master server with the incoming live videos information. The predictive model will predict the number of viewers expected near each cloud site. Based on the predicted results, the optimizer will allocate live videos replicas across the geo-distributed cloud sites near the viewers proximity to minimize the delay and video stalls with the minimum possible cost. Moreover, the optimizer determines from which cloud site the viewers should be served. 
In our work, we consider only the storage resources, while the computation resources for video transcoding are out of the scope of this paper.
\subsection{Predicting live video viewers}\label{section:prediction}
\subsubsection{Dataset}
\begin{figure}[h]
 \captionsetup{font=scriptsize}
  \includegraphics[width=\linewidth]{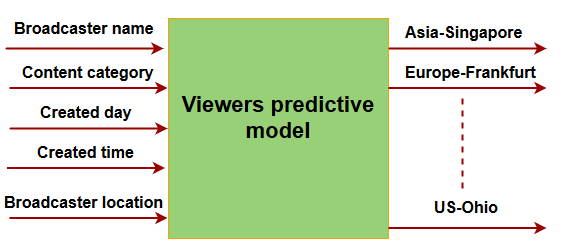}
  \caption{Predictive model input and output.}
  \label{fig:predictive model}
\end{figure}

in our work, we are using the Facebook 2018 live videos dataset collected by our team \cite{facebookvideoslive18}, containing more than two million Facebook live video streams. The active video streams metadata are fetched every 3 minutes in different periods on January, February, March, May, June and July 2018. As a result, we obtained a list of fetches related to each video and containing the number of viewers at the recording time. The live videos are collected with many features such as creation time and day, broadcaster location, number of likes and most importantly the viewers' locations. In this work, we selected six features for each video namely, the broadcaster name, content category, created time, created day, broadcaster location and the viewers' locations as illustrated in Fig. \ref{fig:predictive model}. The viewers' locations were selected from the video fetch with maximum number of viewers.
\begin{figure*}[h]
 \captionsetup{font=scriptsize}
  \includegraphics[height=4cm,width=\linewidth]{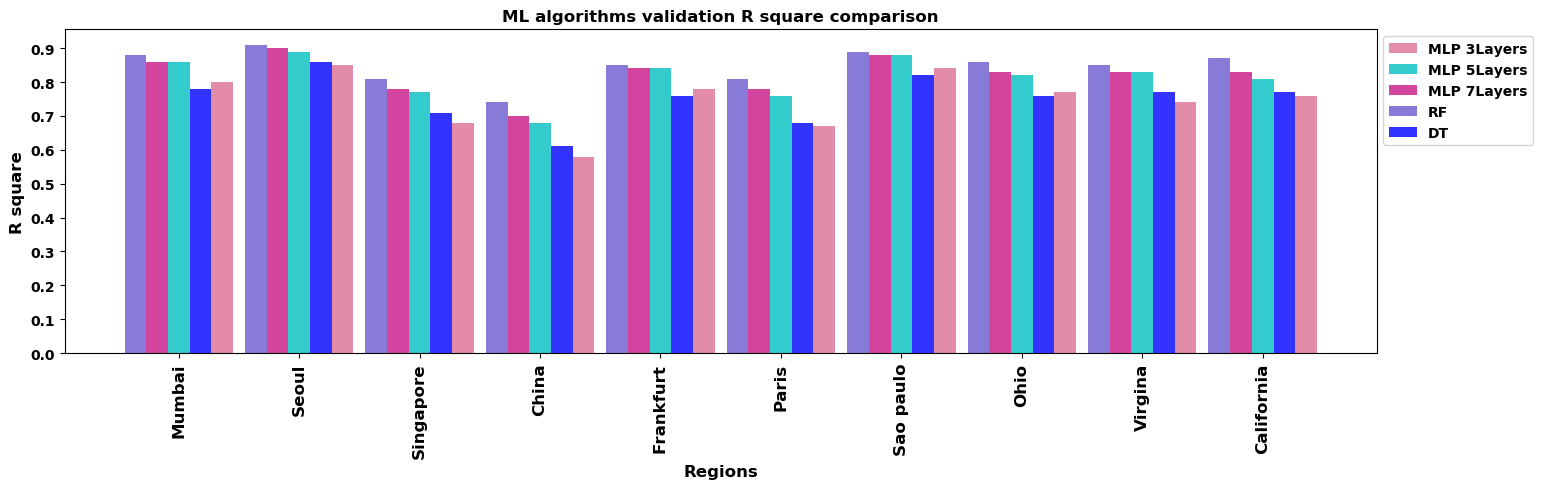}
  \caption{Models validation.}
  \label{fig:Validation}
\end{figure*}
\begin{figure*}[h]
\captionsetup{font=scriptsize}
  \includegraphics[height=4cm,width=\linewidth]{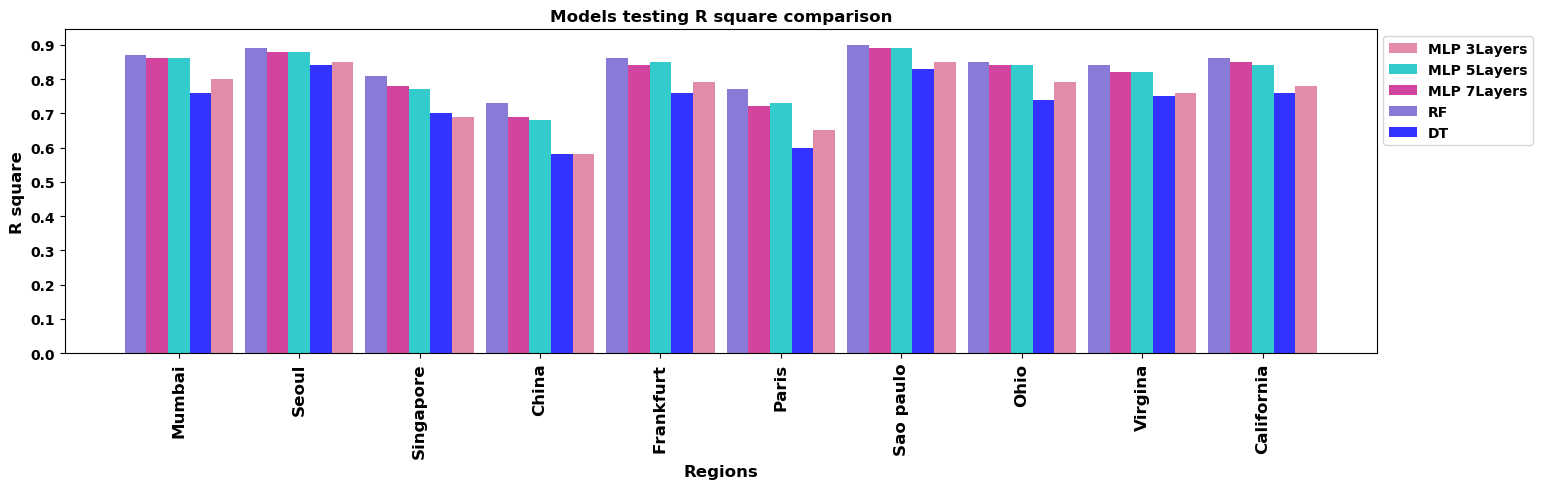}
  \caption{Models testing.}
  \label{fig:Testing}
\end{figure*}

\begin{figure*}[t]
  \centering
   \captionsetup{font=scriptsize}

  \subcaptionbox{\scriptsize{Asia Seoul.}}[.3\linewidth][c]{%
    \includegraphics[scale=0.38]{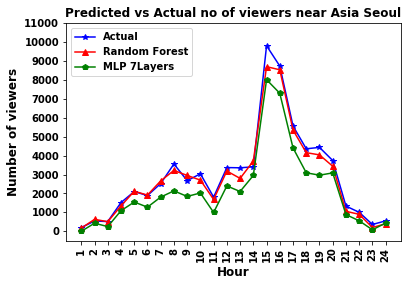}}\quad
  \subcaptionbox{\scriptsize{Europe Frankfurt.}}[.3\linewidth][c]{%
    \includegraphics[scale=0.38]{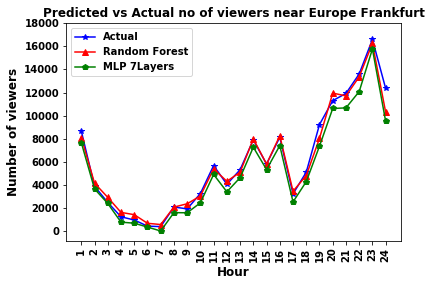}}\quad
  \subcaptionbox{\scriptsize{China Ningxia.}}[.3\linewidth][c]{%
    \includegraphics[scale=0.38]{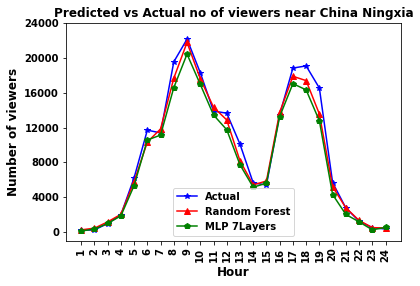}}
  \caption{Hourly actual vs predicted viewers number.}
  \label{fig:predVsActual}
\end{figure*}
\subsubsection{Preprocessing}
as our objective is to predict the viewers number near various geo-distributed cloud sites, there was a need to preprocess our raw data. First, we mapped the viewers’ locations into 10 Amazon Web Services (AWS) cloud sites locations \cite{amazon} namely, Asia-Mumbai, Asia-Seoul, Asia-Singapore, China-Ninxgia, Europe-Frankfurt, Europe-Paris, South America-Sao paulo, US East-Ohio, US East-Virginia and US West-California. This was done by calculating the shortest distance between the viewer’s locations and the 10 AWS cloud sites locations. Furthermore, we calculated the number of viewers near each cloud site for each video. We did the same to the broadcaster location, where we mapped his location into the nearest AWS cloud site.
Moreover, we clustered the created time into 6 time periods. Finally, we applied the categorical one-hot encoding to the time period, created day and broadcaster location features, while we used feature hashing introduced by \cite{weinberger2009feature} to transform the high-cardinality features namely broadcaster name and content category into hashed feature vectors.

\subsubsection{Predictive model}
the dataset used to train our models included 224,839 live video records collected in March, May and June 2018. 80\% of the records were randomly selected for training and 20\% were used for validation.
We trained our regression models to produce 10 outputs as illustrated in Fig. \ref{fig:predictive model}, each represents the number of viewers near the 10 AWS cloud sites mentioned previously. We adopted three different ML algorithms namely, Multilayer-perceptron (MLP), Decision trees (DT) and Random Forest (RF). 
We built several models using each ML algorithm, as there is no method to predetermine the best combination of hyperparameters, such as the number of hidden layers and neurons for MLP models, number of forests for RF models and the max depth for DT models. Finally, the best models were selected considering the best determination coefficient ($R^2$) values,  which is used to assess the goodness of fit of our regression models. $R^{2}$ values approaching 1 indicate that the model provides accurate predictions, and it is calculated according to Eq. (\ref{eq:2}):
\begin{equation}\label{eq:2}
R^{2}=1-\frac{\sum_{i=1}^{m}\left( A_{i}-P_{i}\right)^{2}}{\sum_{i=1}^{m}\left( A_{i}-\bar{A}\right )^{2}}\end{equation}
where m is the number of videos, $A_{i}$ is the actual number of viewers for video $i$, $P_{i}$ is the predicted number of viewers for video $i$, and $\bar{A}$ is the mean of the actual number of viewers of all videos.

\subsubsection{Predictive model results}
after training the models, the validation results, depicted in Fig. \ref{fig:Validation}, showed that RF outperforms the other ML algorithms by achieving for example an $R^2$ of 0.91 for Seoul, 0.89 for Sao Paulo, 0.85 for Ohio, 0.86 for California and 0.74 for China. The DT model achieved the lowest $R^2$ as opposed to MLP and RF. The results showed that increasing the number of layers for the MLP models improves the results. However, due to the complexity of the models, and because we noticed that there is a slight difference between the performance of the 5 layers model and the 7 layers model, we did not increase the layers above 7.
The results also showed that for all ML models, the predicted number of viewers near some regions achieved a higher $R^2$ compared to other regions, China achieved the lowest, while Seoul and Sao paulo achieved the best $R^2$ . We further tested our models on unseen data of live videos collected from July 1 to July 6, 2018. The models performed the same as with validation data in some regions, slightly less or higher in other regions as shown in Fig. \ref{fig:Testing}. We then extended our experiments by performing the predictions on hourly basis for 24 hours using the live videos of July 3, 2018. The RF and MLP 7 layers models were used for prediction, since they performed better than other models. The predicted number of viewers for the hourly incoming live videos versus the actual number of viewers for Seoul, Frankfurt and China cloud sites are presented in Fig. \ref{fig:predVsActual}. Since our results demonstrate that the RF predictions are the closest to the actual values, we will adopt this model in our system.
\subsection{Proactive live video allocation and viewers serving}
In this section, we formulate the problem of proactive resource allocation,  to derive the optimal number of video allocation
cloud sites and the nearest cloud site to serve the viewers, with an objective of minimizing the cost constrained by the access delay. We then, present our proactive resource allocation algorithm. 

\subsubsection{Problem formulation}
the set of incoming live videos at period $t$ is denoted by $V(t)$=\{$v_{1}$, $v_{2}$, $v_{3}$,....$v_{m}$\}. The set of regions is represented by $R$=\{$r_{1}$, $r_{2}$, $r_{3}$,....$r_{n}$\}. Let $r^b$, $r^a$ and $r^w$ denote the broadcasting region, video allocation region and video serving region respectively. The round trip delay from $r^a$ to $r^w$ is represented by $d_{r^{a}r^{w}}$. 
Let $P(t)=\{P_{v_{1}},P_{v_{2}},...P_{v_{m}}\}$ represent the set of predicted viewers for the incoming videos at period $t$. As each video has predicted viewers in different regions, 
let $P_v$=\{$p_1$, $p_2$, $p_3$,....$p_n$\} denote the set of the number of predicted viewers at different regions for each video $v$. The broadcasters' regions for the incoming videos at period $t$ is denoted by $B(t)=\{r^{b}_1,r^{b}_2,...r^{b}_m\}$.
Due to the fact that some videos do not have any viewers near some cloud sites, let $E(v, r^{w})$ present a binary variable, equal to 1, if video $v$ has predicted viewers near the region $r^{w}$, and 0 otherwise. 

We consider renting S3 storage \cite{amazonS3} servers at each cloud site. Three types of costs are taken into account: (1) the storage cost at each cloud site; (2) the migration cost of a video replica from one cloud site to another and (3) the cost of serving viewers. We assume that the storage capacity can be provisioned based on the application demand. On allocation cloud site at region ${r^{a}}$, let $\alpha_{r^{a}}$ be the storage cost per GB, which varies based on site location and the storage thresholds fixed by Amazon S3. For example, Amazon charges 0.023\$ per GB for the first 50TB, while it charges 0.021\$ when exceeding 500TB in the case of US East Virginia region \cite{amazonS3}.
Given that $\kappa$ is the video size, the total storage cost $\mathbb{S}$ can be calculated as presented in Eq. \ref{eq:S}.
Given that $\eta_{r^{b}}$ is the cost to migrate a copy of a video from the broadcaster region ${r^{b}}$ to allocation region ${r^{a}}$, which is the data transfer cost from one cloud site to another per GB, the total migration cost $\mathbb{M}$ is calculated as presented in Eq. \ref{eq:M}. 
Given that $\omega_{r^{a}}$ is the serving request cost from region ${r^{a}}$, which is the data transfer cost from that region to the internet per GB, the total serving request cost $\mathbb{R}$ is calculated as presented in Eq. \ref{eq:R}. The overall cost $\mathbb{C}$ to serve viewers is shown in Eq. \ref{eq:Cost}.\\
$p_{r^{w}}$ is the predicted number of viewers at region $r^w$. 
\begin{equation}\label{eq:S}
\begin{aligned}
\mathbb{S}=\sum_{v\in V(t)}\sum_{r^{a}\in R}\alpha _{r^{a}}*\kappa *A(v,r^{a})
\end{aligned}
\end{equation}
\begin{equation}\label{eq:M}
    \begin{aligned}
    \mathbb{M}=\sum_{v\in V(t)}\sum_{r^{a}\in R}\eta _{r^{b}}*\kappa *A(v,r^{a})
    \end{aligned}
\end{equation}
\begin{equation}\label{eq:R}
    \begin{aligned}
    \mathbb{R}=\sum_{v\in V(t)}\sum_{r^{a}\in R}\sum_{r^{w}\in R}\omega_{r^{a}}*\kappa*p_{r^{w}}*W(v,r^{a},r^{w})
    \end{aligned}
\end{equation}
\begin{equation}\label{eq:Cost}
\begin{aligned}
\mathbb{C}=\mathbb{S}+ \mathbb{M} + \mathbb{R}
\end{aligned}
\end{equation}
\begin{subequations}\label{first:main}
Our objective is to minimize the cost for period t as shown in Eq. \ref{first:main}:
\begin{equation}
\min_{A(v, r^{a}) \ W(v, r^{a},r^{w})}  \mathbb{C}
\tag{\ref{first:main}}
\end{equation}
Subject to the following constraints:

Every video is allocated by default in the broadcaster nearest cloud site.
\begin{equation}
A(v, r^{b})=1   \hspace{4em}    \forall v \in V(t), \forall r^{b} \in B(t)\label{first:a}
\end{equation}

A video $v$ can be served from region $r^{a}$ to viewers at region $r^{w}$, only if it is allocated at region $r^{a}$.
\begin{equation}
 W(v, r^{a}, r^{w})\leq A(v, r^{a})
 \hspace{2em}\forall v \in V(t), \forall r^{a} \in R,\forall r^{w} \in R \label{first:b}
\end{equation}

A video $v$ can be served from region $r^{a}$ to $r^{w}$ only if there exists viewers at $r^{w}$.
\begin{equation}
W(v, r^{a}, r^{w})\leq E(v, r^{w})  \hspace{2em}\forall v \in V(t), \forall  r^{a} \in R,\forall  r^{w} \in R\label{first:c}
\end{equation}

If there exists viewers for video $v$ at region $r^{w}$, they can only be served from one region.  

\begin{align}
\sum_{r^{a}\in R}W(v, r^{a}, r^{w})=E(v, r^{w}) \hspace{1em} \forall v \in V(t), \forall r^{w} \in R \end{align}\label{first:d}

The average serving request delay for each video should not exceed a threshold $\mathbb{D}$.

\begin{align}
\frac{\sum_{r^{a}\in R}\sum_{r^{w}\in R} p_{r^{w}}*d_{r^{a}r^{w}}*W(v,r^{a},r^{w})}
{\sum_{r^{w}\in R}p_{r^{w}}} \leq \mathbb{D} \hspace{1em}\forall v \in V(t)\end{align}\label{first:e}

Binary decision variables that can be set to 0 or 1.
\begin{align}
A(v, r^{a}), W(v, r^{a}, r^{w}) \in \{0,1\}\end{align}\label{first:f}

\end{subequations}
The decision variable $A(v,r^a)$ is equal to 1, if video $v$ is allocated in region $r^a$, and 0 otherwise. While the decision variable $W(v,r^a,r^w)$ is equal to 1, if viewers at region $r^w$ are served from region $r^a$ and 0 otherwise. The problem formulation notations are presented in Table  \ref{table:notations}. 
\subsubsection{Proactive resource allocation}
the proposed proactive resource allocation algorithm is presented in Algorithm \ref{Proactive}. In fact, at each period t, the system receives a set of incoming videos, which will be an input to the viewers predictive model. Based on the predicted viewers, the optimal number of allocation cloud sites and the nearest cloud site to serve the viewers will be decided by the optimizer. The storage resources at each cloud site is reserved based on the allocation decisions, and released for ended live videos from the previous periods. Moreover, the viewers are served from their closest cloud site based on the serving decisions.

\begin {table}[h]

\captionsetup{font=scriptsize}
\caption {Notations for the formalized problem.}
\label{table:notations}
\centering
\begin {tabular}{ l p{6 cm}  }

\hline
 \textbf{Notation} &  \textbf{Description}   \\ [0.5ex] 
 \hline
 $V(t)$ & Set of incoming live videos at period $t$  \\ 
 $R$ & Set of regions  \\ 
 $B(t)$ & Set of broadcasters regions for videos at period $t$  \\ 
 $r^{a}$ & Region of video allocation \\ 
$r^{w}$ & Region of serving \\
$r^{b}$ & Region of broadcasting  \\
$P(t)$ &Set of predicted viewers for live videos at period $t$\\
$P_v$ & Set of predicted viewers at different $R$ for video $v$  \\
$SU$& Set of storage used at each region\\
$W(v, r^{a}, r^{w})$ & Binary decision variable that indicates the serving site\\
$A(v, r^{a})$ & Binary decision variable that indicates the allocation site\\
$E(v, r^{w})$ & Binary variable that indicates viewers existence\\
 $d_{r^{a}r^{w}}$ & Round trip delay between ${r^{a}}$ and ${r^{w}}$ \\
 $RTT$& Matrix for round trip delay between the different $R$\\ 
  $\mathbb{D}$ & Delay threshold\\
  $\kappa$ & Video size\\
 $\alpha_{r^{a}}$ & Storage cost per GB at region ${r^{a}}$ \\
  $\eta_{r^{b}}$& Migration cost per GB from broadcaster region ${r^{b}}$ \\
 $\omega_{r^{a}}$ & Serving request cost per GB from ${r^{a}}$ \\
 $\mathbb{S}$ & Total storage cost\\
 $\mathbb{M}$ & Total migration cost\\
 $\mathbb{R}$ & Total serving request cost \\
 $\mathbb{C}$ & Overall cost \\
 \hline
\end {tabular}
\end {table}
\begin{algorithm}[!h]
\caption {Proactive resources allocation}
\label{Proactive}
\begin{algorithmic}[1]
\State \footnotesize{\textbf{Input: $R$, \{$\alpha_1,...,\alpha_n$\}, 
\{$\eta_1,...,\eta_n$\}, \{$\omega_1,...,\omega_n$\}, $RTT$, $\kappa$
}}
\State \footnotesize{\textbf{Storage usage at each cloud site initialization: }$SU_1=0,..., SU_n=0$}
\For {$t \in{\{1,..,T\}}$}
\State \footnotesize{- Receive videos informations $V(t)$ and their broadcasters $B(t)$}
\State \footnotesize{- Run predictive model to predict $P(t)$ for videos $V(t)$}
\State \footnotesize{- Derive optimal solution $\min_{A(v,r^a)W(v,r^a,r^w)}$ $\mathbb{C}$ as per Eq. \ref{first:main}}

\For {$j \in {\{1,..,n\}}$} 
\State \footnotesize{- Update $SU_j$ based on allocation decisions $A(v,r^a)$}
\State \footnotesize{- Release storage resources for ended videos from previous}
\State\footnotesize{ periods.}
\State \footnotesize{- Serve viewers based on serving decisions $W(v,r^a,r^w)$}
\EndFor
\EndFor
\end{algorithmic}
\end{algorithm}
\section{Performance Evaluation}\label{section:performance}
\subsection{Simulation settings}
\begin{figure}[h]
\captionsetup{font=scriptsize}
\centering
  \includegraphics[height=4cm,width=7cm]{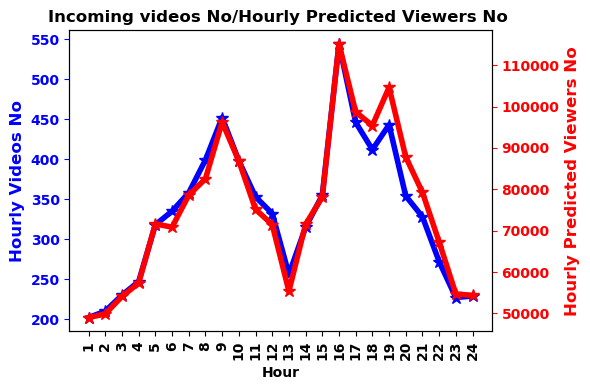}
  \caption{Hourly incoming videos/ Hourly predicted viewers.}
  \label{fig:ViewersNo/VideoNo}
\end{figure}

In this section, we evaluate the performance of our system using the RF hourly predicted viewers of July 3, 2018 to get the hourly optimal resource allocation for $T$=24(hours) and $t$=1(hour). The number of hourly incoming videos, and the hourly predicted viewers used in our simulation are presented in Fig. \ref{fig:ViewersNo/VideoNo}. In our system, we assume that the video duration is 4 hours, which is the maximum video duration for a Facebook live video. We assume that if a video is allocated in a set of cloud sites at period $t$, it will be allocated in the same cloud sites for the remaining time periods of streaming. Moreover, because video quality is out of the scope of this paper we assume that the viewers are served with the best video quality, where we set the video size $\kappa$ to 0.738 Gbit.
We constructed our round trip time (RTT) matrix $d_{r^{a}r^{w}}$ by calculating the average RTT from the different cloud sites using \cite{wondernetwork} accessed on September 19, 2018. The storage and data transfer prices of Amazon S3 \cite{amazonS3} are considered in our simulation to model $\alpha$, $\omega$ and $\eta$. We varied the latency thresholds constraints $\mathbb{D}$ for serving a video to 8.8ms, 60ms, 120ms, 171ms, 220ms and 371ms. 8.8ms is the latency needed to serve a viewer from its closest cloud region \cite{bilal2018qoe}.
\begin{figure*}[htb]
\captionsetup{font=scriptsize}
  \centering
  \subcaptionbox{\scriptsize{Hourly optimal cost.}}[.3\linewidth][c]{%
    \includegraphics[width=.35\linewidth]{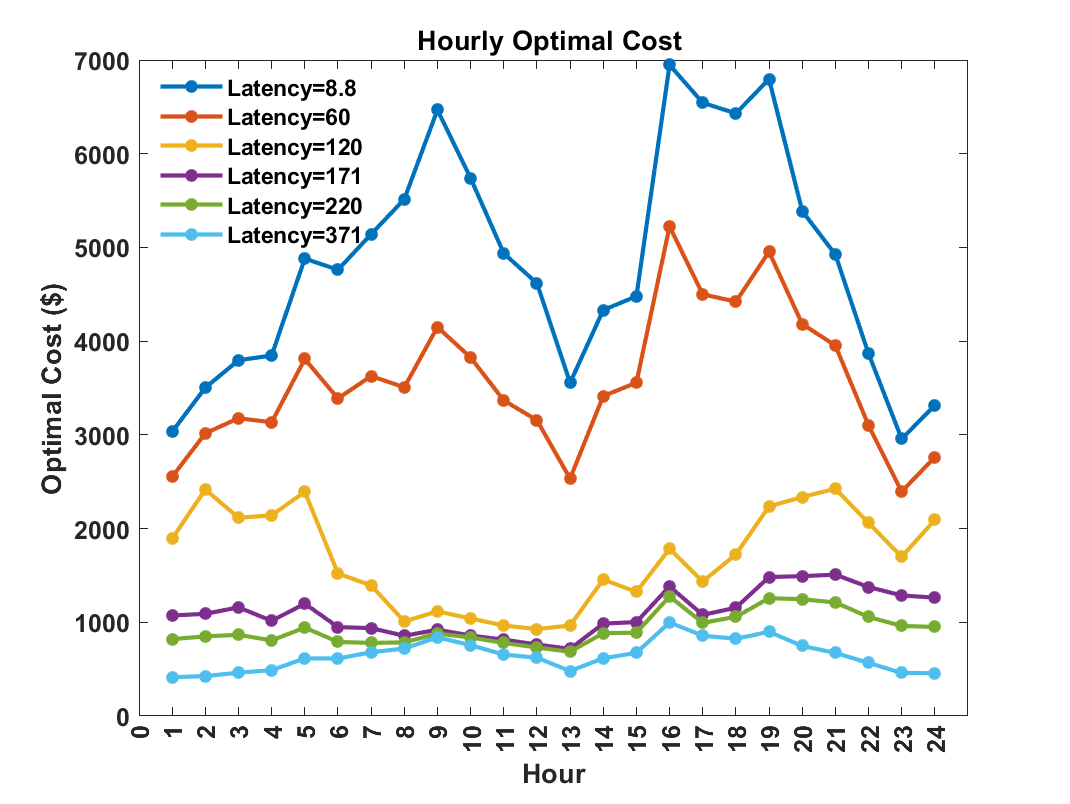}}\quad
  \subcaptionbox{\scriptsize{Total cost vs latency thresholds.}}[.3\linewidth][c]{%
    \includegraphics[width=.35\linewidth]{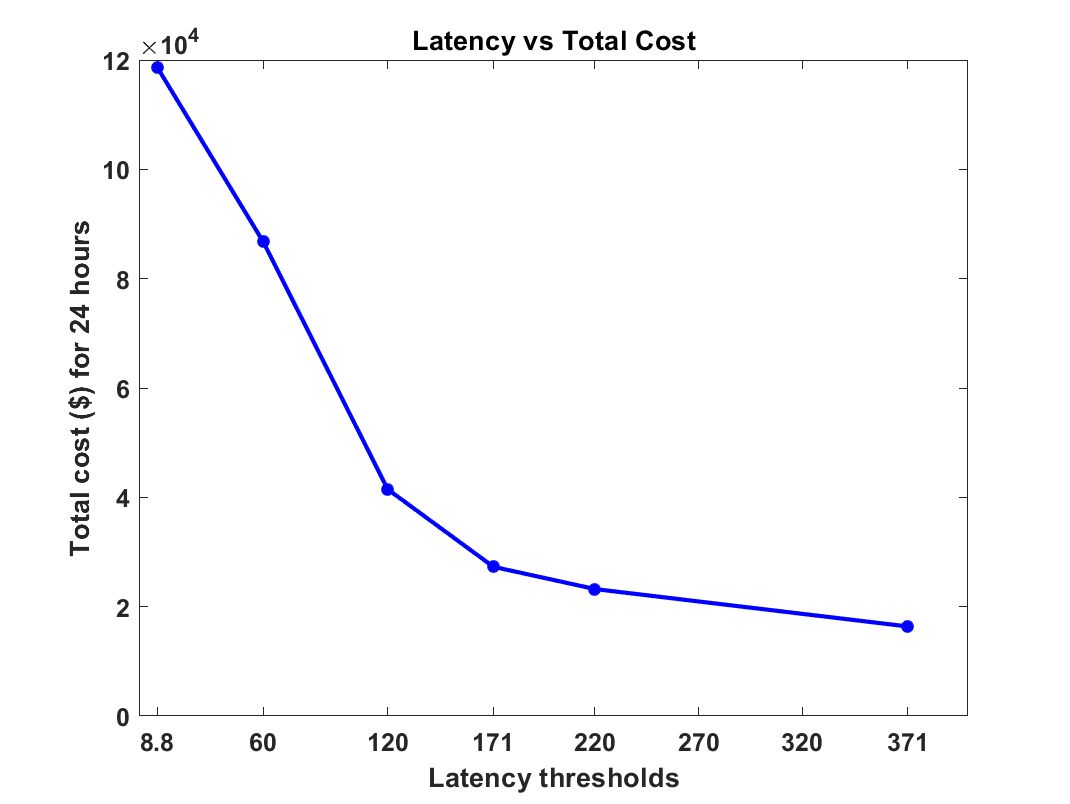}}\quad
  \subcaptionbox{\scriptsize{Serving hits percentages.}}[.3\linewidth][c]{%
    \includegraphics[width=.35\linewidth]{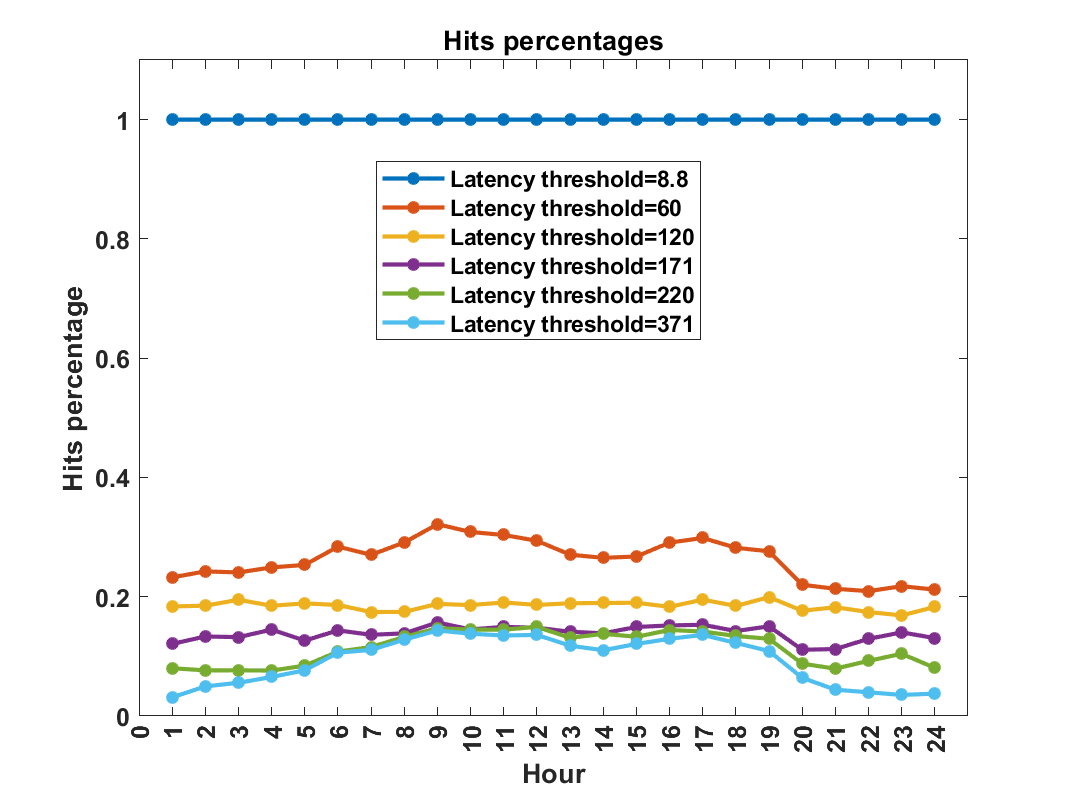}}
  \caption{Simulation results.}
  \label{fig:simulationResults}
\end{figure*}
\begin{figure}[h]
\captionsetup{font=scriptsize}
\centering
\includegraphics[height=8cm,width=9cm]{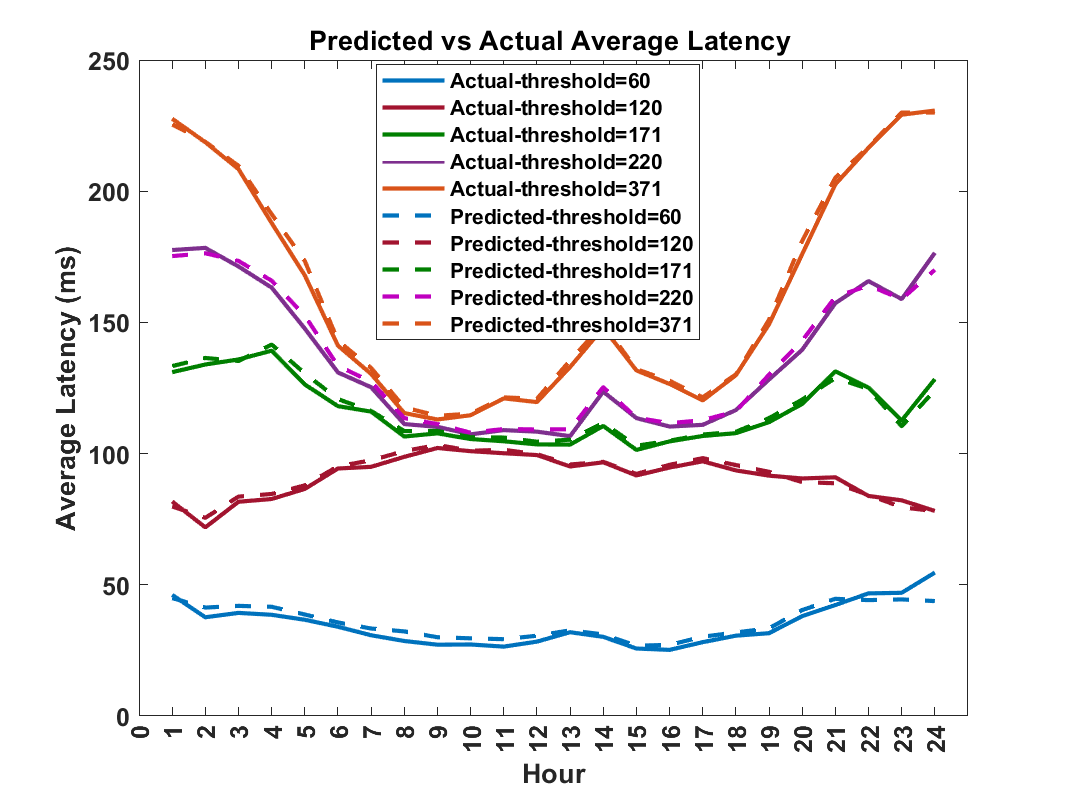}
  \caption{Predicted vs actual hourly average latency.}
  \label{fig:latencyActualvsPredicted}
\end{figure}
\subsection{Simulation results}
Fig. \ref{fig:simulationResults}(a) shows that we can establish a trade-off between the video access delay and the resource allocation cost. Indeed, the hourly optimal cost is high when the system is forced to serve the viewers from their region by setting the latency threshold to 8.8ms. Relaxing the threshold leads to minimizing the cost. Therefore, the content provider can sacrifice in terms of cost to enhance the QoE or the opposite based on his requirements. It is worth mentioning that the optimal cost is higher in some periods as opposed to others, because as illustrated in Fig. \ref{fig:ViewersNo/VideoNo} the number of incoming videos and predicted viewers varies from period to another.

In order to evaluate the total system cost over the 24 hours with various latency thresholds, we calculated the hourly total cost, as presented in Fig. \ref{fig:simulationResults}(b). The hourly total cost is defined as the sum of the network cost at period t and the cost of storage of still running videos, which is presented in Eq. \ref{eq:TH}, given that $SU_{n}$ is the storage usage at region $n$ until period $t$.
\begin{equation}\label{eq:TH}
\begin{aligned}
\text{Hourly total cost (t)}=\mathbb{C}(t)+\sum_{r^{a}\in R} \alpha_{r^{a}}*SU_{r^a}
 \end{aligned}
\end{equation}
The system total cost is calculated as shown in Eq. \ref{eq:T}:
\begin{equation}\label{eq:T}
\begin{aligned}
\text{System total cost}=\sum_{t=1}^{T} \text{Hourly total cost(t)}
\end{aligned}
\end{equation}

Furthermore, we calculated the hits percentages, which represents the percentage of videos served from the same region of viewers as shown in Fig. \ref{fig:simulationResults}(c). Setting the latency to 8.8ms resulted in hits percentage of 100\% in every hour, as all viewers will be served from their region. While it is in the range of 20\% to 30\% with 60ms latency threshold. Moreover, when the latency threshold was set to 120ms, 171ms, 220ms and 371ms, less than 20\% of videos were served from the same region of viewers. The hits percentage was very low with high latency thresholds, as the system is not forced to serve the viewers from their closest region. 

Finally, to evaluate the accuracy of our resource allocation framework, we calculated the hourly average latency using the proactive serving decisions with variant latency thresholds $\mathbb{D}$. In fact, we calculated the latency of serving the actual number of viewers based on our proactive video allocation and we compared it to the latency derived from the predictive model. The results as shown in Fig. \ref{fig:latencyActualvsPredicted} proved that the average latency to serve the actual viewers is very close to the average latency serving the predicted viewers. Moreover, the average latency to serve the actual viewers did not exceed the latency thresholds $\mathbb{D}$. 
\section{Conclusion}\label{section:conclusion}
In this paper, we propose a proactive resource allocation framework. First, we adopt machine learning to build a predictive model that captures the viewers number near each geo-ditributed cloud site.
Then, based on the predicted results, we formulated our resource allocation model as an optimization problem to optimally allocate resources across the geo-distributed cloud sites based on the viewers proximity. 
For the future work, we plan to design a distributed proactive resource allocation framework. We are also interested in implementing predictive models for the number of incoming live videos, the live video duration, the live videos viewing time and the computation resources. 
\section*{Acknowledgment}
This publication was made possible by NPRP grant 8-519-1-108 from the Qatar National Research Fund (a member of Qatar Foundation). The findings achieved herein are solely the responsibility of the author(s)
\printbibliography
\end{document}